\documentstyle[]{article}
\title{ Vortices and hierarchy of states in double-layer
        fractional Hall effect }
\author{Jacek Dziarmaga  \\
        Jagellonian University, Institute of Physics,  \\
        Reymonta 4,30-059 Krak\'ow, Poland
        \thanks{ E-mail address: ufjacekd@ztc386a.if.uj.edu.pl}\\}

\date{21 July 1994}
\begin{document}
\maketitle

    \begin{abstract}
        Vortex solutions in $U(1)\times U(1)$ Chern-Simons theory
  coupled to a pair of hard-core bosons representing two
  layers of electrons are analysed. It is shown that there is such a range
  of parameters $(\alpha\beta<\gamma^{2})$ in which finite charge $(1,0)$
  or $(0,1)$ vortices can not exist. Instead the minimal
  flux quanta are $(1,1)$ hybrids. Their statistical dominance is
  shown to have an influence on Haldane-Halderin hierarchy of states.
  The form of Coulomb potential and its influence
  on the range of physically acceptable states is discussed.
    \end{abstract}

\section{Introduction}

    The two-dimensional electron system can have a rich structure.
One of the most significant discoveries is the fractional Hall effect
in strong magnetic fields \cite{hall}. The standard theoretical approach
is through variational analysis with trial functions \cite{laughlin}.
There are also field-theoretical studies of the subject based
on generelisations of Ginzburg-Landau model with a use of effective
Chern-Simons
theory \cite{ginzland}. More microscopical treatment of the phenomenon
was introduced by Ezawa and Iwazaki \cite{ei1}. Recently there was discovered
the even-denominator filling factor in double-layer electron system \cite{2l}.
The system was proposed to be described by $U(1)\times U(1)$ Chern-Simons
theory
coupled to a pair of hard-core bosons representing two layers of electrons
\cite{ei2}. Nonperturbative excitations are topological vortices with
some winding numbers on both of the layers $(p,q)$. If $(1,0)$ and
$(0,1)$ vortices condense the hierarchy of states was shown \cite{ei2}
to begin by
\begin{equation}\label{in10}
   \nu^{(1)}=\frac{\nu^{(0)}}{1+\frac{\nu^{(0)}}{4q}} \;\;,
\end{equation}
with characteristic $4q$ instead of $2q$ as in a single-layer system
($q$ is an integer). By carefull analysis we show that there is such
a range of statistical parameters in which minimal vortex excitations with
finite charge are $(1,1)$ vortices living on both of the layers. In this
respect this work is a generalisation of our previous paper \cite{only}.
The physical consequence of this fact is that the denominator is once
again $2q$.

   Another question is the form of Coulomb potential. To have finite
Coulomb energy of the condensates one needs regularising background.
If the background charge density is just opposite to the condensate
density on a given layer the whole spectrum of states $(k,l,m)$ is
allowed. If instead the background is uniform then only states with
$k=l$ survive switching on the Coulomb interaction.

\section{ The model, condensates and vortex solutions }

    The model we consider in this paper was introduced by Ezawa and Iwazaki
\cite{ei2}. It consists of two species of polarised electrons represented by
hard-core bosons coupled
to $U(1)\times U(1)$ Chern-Simons field the effect of which is via
singular gauge transformation equivalent to imposing the Fermi-Dirac
statistics.
\begin{eqnarray}\label{mcv10}
  L&=&i\psi^{\star}D_{0}\psi-\frac{1}{2M}D_{k}\psi^{\star}D_{k}\psi
   +i\phi^{\star}D_{0}\phi-\frac{1}{2M}D_{k}\phi^{\star}D_{k}\phi
   -V(\psi,\phi)                                               \nonumber \\
   &-&\frac{1}{4X}\varepsilon^{\mu\nu\lambda}
                  a_{\mu}^{(1)}\partial_{\nu}a_{\lambda}^{(1)}
     -\frac{1}{4Y}\varepsilon^{\mu\nu\lambda}
                  a_{\mu}^{(2)}\partial_{\nu}a_{\lambda}^{(2)} \nonumber \\
   &-&\frac{1}{4Z}\varepsilon^{\mu\nu\lambda}
                  a_{\mu}^{(1)}\partial_{\nu}a_{\lambda}^{(2)}
     -\frac{1}{4Z}\varepsilon^{\mu\nu\lambda}
                  a_{\mu}^{(2)}\partial_{\nu}a_{\lambda}^{(1)}
     +V(\psi^{\star}\psi,\phi^{\star}\phi)+\triangle V^{C}  \;\;.
\end{eqnarray}
The coefficients $X,Y,Z$ are defined in an entangled way by another
set of parameters
\begin{equation}\label{mcv20}
   X=\frac{\alpha\beta-\gamma^{2}}{\beta}  \;\;,\;\;
   Y=\frac{\alpha\beta-\gamma^{2}}{\alpha} \;\;,\;\;
   Z=\frac{\gamma^{2}-\alpha\beta}{\gamma} \;\;.
\end{equation}
These parameters take values $\alpha=\pi k,\beta=\pi l, \gamma=\pi m$ with
$k,l,m$ odd integers which provide electrons $\psi,\phi$ with desired
statistics. The potential term has the form characteristic for
bosonic end perturbation theory with respect to the parameters
$\alpha,\beta,\gamma$, see \cite{ei1} and e.g.\cite{amelino}
\begin{equation}\label{mcv25}
  V=\frac{1}{2M}[\alpha\mid\psi\mid^{4}
                +2\gamma\mid\psi\mid^{2}\mid\phi\mid^{2}
                +\beta\mid\phi\mid^{4}] \;\;,
\end{equation}
This potential is a field-theoretical remnant of the two-particle
delta-type potential \cite{amelino}. Another potential term is a nonlocal
Coulomb interaction
\begin{equation}\label{mcv26}
  \triangle V^{C}=
   \triangle V^{C}_{11}+\triangle V^{C}_{12}+\triangle V^{C}_{22} \;\;.
\end{equation}
The particular terms are given by
\begin{eqnarray}\label{mcv27}
  \triangle V^{C}_{11}&=&\frac{e^{2}}{2\varepsilon}\int d^{2}x\; d^{2}y  \;
          [\psi^{\star}\psi(x)-\rho^{0}_{1}] \;
          \frac{1}{\mid x-y\mid}             \;
          [\psi^{\star}\psi(y)-\rho^{0}_{1}]                \;\;,\nonumber\\
  \triangle V^{C}_{12}&=&\frac{e^{2}}{\varepsilon}\int d^{2}x\; d^{2}y   \;
          [\psi^{\star}\psi(x)-\rho^{0}_{1}] \;
          \frac{1}{\sqrt{(x-y)^{2}+d^{2}}}   \;
          [\phi^{\star}\phi(y)-\rho^{0}_{2}]                \;\;,\nonumber\\
  \triangle V^{C}_{22}&=&\frac{e^{2}}{2\varepsilon}\int d^{2}x\; d^{2}y  \;
          [\phi^{\star}\phi(x)-\rho^{0}_{2}] \;
          \frac{1}{\mid x-y\mid}             \;
          [\phi^{\star}\phi(y)-\rho^{0}_{2}]                  \;\;,
\end{eqnarray}
where $\varepsilon$ is a dielectric constant and $d$ is the interlayer
distance. We have taken into account uniform background the charge density
of which is opposite to the medium charge density of electrons.
The Lagrangian in this form does make sense only if
$\alpha\beta-\gamma^{2}\neq 0$ - the case we refer to as nondegenerate,
but the model itself can be made more general.
Couplings between electrons and gauge fields are established by
\begin{equation}\label{mcv30}
  D_{\mu}\psi=\partial_{\mu}\psi-ia_{\mu}^{(1)}\psi+ieA_{\mu}^{ext}\psi
\;\;,\;\;
  D_{\mu}\phi=\partial_{\mu}\phi-ia_{\mu}^{(2)}\phi+ieA_{\mu}^{ext}\phi \;.
\end{equation}
In what follows we will consider only external magnetic field and
put $A^{ext}_{0}=0$. Variation with respect to Lagrange multipliers
$a_{0}^{(I)},\;I=1,2$ leads to constraints
\begin{eqnarray}\label{mcv40}
  B_{1}&=&2\alpha\rho_{1}+2\gamma\rho_{2}          \;\;, \nonumber \\
  B_{2}&=&2\gamma\rho_{1}+2\beta\rho_{2}           \;\;,
\end{eqnarray}
where the electronic densities are
$\rho_{1}=\psi^{\star}\psi \;,\; \rho_{2}=\phi^{\star}\phi$. The Hamiltonian
of the model is
\begin{equation}\label{mcv50}
  H=\frac{1}{2M}D_{k}\psi^{\star}D_{k}\psi
   +\frac{1}{2M}D_{k}\phi^{\star}D_{k}\phi+V+\triangle V^{C}
\end{equation}
For the following we will switch off the Coulomb interaction and later on
regard it as a correction.
In a uniform external magnetic field Bogomol'nyi decomposition
\cite{bogomol} leads to
\begin{eqnarray}\label{mcv60}
  H=\int d^{2}x \; [ \; \frac{1}{2M}\mid(D_{1}+iD_{2})\psi\mid^{2}
   +\frac{1}{2M}\mid(D_{1}+iD_{2})\phi\mid^{2}
   +\frac{eB^{ext}}{2M}(\rho_{1}+\rho_{2})            \nonumber \\
   +V-\frac{1}{2M}(\rho_{1}B_{1}+\rho_{2}B_{2})
   -\nabla\times\vec{J}_{1}-\nabla\times\vec{J}_{2} \; ] \;.
\end{eqnarray}
If we neglect boundary terms and take into account Gauss' laws
(\ref{mcv40}) the second line will
vanish identically. The term $\rho_{1}+\rho_{2}$ in the first line
can be integrated out if we impose an extra costraint of definite number
of particles $N$ \cite{ei1,ei2}.
Thus we are left with the Hamiltonian
\begin{equation}\label{mcv70}
  H=\frac{1}{2}\omega_{c}N
   + \int d^{2}x \; [ \; \frac{1}{2M}\mid D_{+}\psi\mid^{2}
   +\frac{1}{2M}\mid D_{+}\phi\mid^{2} \; ] \;\;,
\end{equation}
where $\omega_{c}=\frac{eB^{ext}}{M}$ is a cyclotron frequency.
Minimal energy static solutions satisfy self-dual equations
\begin{equation}\label{mcv80}
  D_{+}\psi\equiv\partial_{+}\psi-ia_{+}^{(1)}\psi+ieA^{ext}_{+}\psi=0
  \;\;,\;\; D_{+}\phi=0 \;\;,
\end{equation}
in addition to Gauss' laws (\ref{mcv40}). Solutions to these equations
are also solutions of full Euler-Lagrange equations of the model only
if Lagrange multiplyers are equal to
\begin{equation}\label{mcv85}
  a_{0}^{(1)}=-\alpha\rho_{1}-\gamma\rho_{2} \;\;\;,\;\;\;
  a_{0}^{(2)}=-\gamma\rho_{1}-\beta\rho_{2} \;\;.
\end{equation}
For a solution with constant densities $\rho_{1},\rho_{2}$ the phases
of scalar fields can be put constant and Eqs.(\ref{mcv80}) imply that
\begin{equation}\label{mcv90}
  B_{1}=eB^{ext} \;\;\;,\;\;\; B_{2}=eB^{ext} \;\;.
\end{equation}
These and the constraints (\ref{mcv40}) lead to the following values
of uniform condensates in the nondegenerate case
\begin{equation}\label{mcv100}
  \rho^{0}_{1}=\frac{1}{2}eB^{ext}\frac{\beta-\gamma}{\alpha\beta-\gamma^{2}}
\;\;\;,\;\;\;
  \rho^{0}_{2}=\frac{1}{2}eB^{ext}\frac{\alpha-\gamma}{\alpha\beta-\gamma^{2}}
\;\;.
\end{equation}
For positive value of $B^{ext}$, which we take for definiteness, the nonzero
condensates exist if and only if $\alpha-\gamma$,$\beta-\gamma$ and
$\alpha\beta-\gamma^{2}$ are simultaneously positive or negative.
The total electronic filling factor in this case is
\begin{equation}\label{mcv110}
  \nu^{(0)}\equiv\frac{2\pi(\rho_{1}+\rho_{2})}{eB^{ext}}=
           \frac{\pi(\alpha+\beta-2\gamma)}{\alpha\beta-\gamma^{2}}=
           \frac{k+l-2m}{kl-m^{2}} \;\;,
\end{equation}
Thus for e.g. $(k,l,m)=(3,3,1)$ we have the experimentally observed
filling factor $\frac{1}{2}$ \cite{2l}.

  In the degenerate case there is a continuous family of constant
solutions characterised by a constraint
\begin{equation}\label{mcv120}
  \rho\equiv\rho_{1}+\rho_{2}=\frac{eB^{ext}}{2\gamma} \;\;,
\end{equation}
only if in addition $\alpha=\beta=\gamma$. The filling factor
is equal to $\nu^{(0)}=\frac{1}{m}$ with $m$ being an odd integer. A free
parameter in the solution signals existence of a zero mode.

  Thanks to the form of Coulomb potential all the uniform solutions
(\ref{mcv27}) have zero electrostatic energy. On the other hand contrary
to the assumption in (\ref{mcv27}) one could think it to be more natural
that the positive charge density of the background is constant everywhere
and in particular the same for the two layers
$\rho^{0}=\frac{1}{2}(\rho^{0}_{1}+\rho^{0}_{2})$. In such a case
the Coulomb potentials should be taken as
\begin{eqnarray}\label{mcv123}
  \triangle \hat{V}^{C}_{11}&=&\frac{e^{2}}{2\varepsilon}\int d^{2}x\; d^{2}y
\;
          [\psi^{\star}\psi(x)-\rho^{0}]        \;
          \frac{1}{\mid x-y\mid}            \;
          [\psi^{\star}\psi(y)-\rho^{0}]                  \;\;,\nonumber\\
  \triangle \hat{V}^{C}_{12}&=&\frac{e^{2}}{\varepsilon}\int d^{2}x\; d^{2}y
\;
          [\psi^{\star}\psi(x)-\rho^{0}]        \;
          \frac{1}{\sqrt{(x-y)^{2}+d^{2}}}  \;
          [\phi^{\star}\phi(y)-\rho^{0}]                  \;\;,\nonumber\\
  \triangle \hat{V}^{C}_{22}&=&\frac{e^{2}}{2\varepsilon}\int d^{2}x\; d^{2}y
\;
          [\phi^{\star}\phi(x)-\rho^{0}]        \;
          \frac{1}{\mid x-y\mid}            \;
          [\phi^{\star}\phi(y)-\rho^{0}]                  \;\;.
\end{eqnarray}
With this background any constant solution such that
$\rho_{1}\neq\rho_{2}$ would have infinite energy.
Thus the only physical states would be those with $\alpha=\beta$ or
in other words one could observe only states of the type $(k,k,m)$,
which characterise by filling factors $\nu^{(0)}=\frac{1}{s}$, where
$s$ is any integer (even or odd).
In the degenerate case there would be no longer any free parameter
but rather the separate charge densities would be locked at
$\rho_{1}=\rho_{2}=\frac{eB^{ext}}{4\gamma}$. Thus the zero mode can
be expected to become massive thanks to Coulomb energy as it
was observed experimentally \cite{2l}.

  Now let us take into account a possibility of nonuniform vortex
solutions. Let us write the scalar fields in the form
$\psi=\sqrt{\rho_{1}}\exp i\omega_{1},
 \phi=\sqrt{\rho_{2}}\exp i\omega_{2}$. The self-dual equations
(\ref{mcv80}) combined together with Gauss' laws (\ref{mcv40})
lead to the following modified Toda equations
\begin{eqnarray}\label{mcv125}
  \frac{1}{2}\nabla^{2}\ln\rho_{1}
      &=&2\alpha\rho_{1}+2\gamma\rho_{2}-eB^{ext}
        +\varepsilon_{mn}\partial_{m}\partial_{n}\omega_{1} \;\;,\nonumber\\
  \frac{1}{2}\nabla^{2}\ln\rho_{2}
      &=&2\gamma\rho_{1}+2\beta\rho_{2}-eB^{ext}
        +\varepsilon_{mn}\partial_{m}\partial_{n}\omega_{2} \;\;,
\end{eqnarray}
The phases $\omega_{I}$ are single-valued except of finite sets of singular
points. Outside of these points one can neglect the last terms in
above equations. It is easy to see that if we impose a condition
$\rho_{2}=\frac{\alpha-\gamma}{\beta-\gamma}\rho_{1}$ inspired by
the uniform solutions (\ref{mcv100}) we will obtain an equation
\begin{equation}\label{mcv130}
  \frac{1}{2}\nabla^{2}\ln\rho_{1}
         =\frac{2(\alpha\beta-\gamma^{2})}{\beta-\gamma}\rho_{1}-eB^{ext} \;\;,
\end{equation}
which is known to possess multivortex solutions. Thus in any case
there are static multivortex solutions with vortices with winding
numbers (1,1) located at arbitrary points of the plane. We will call
them hybrid anyons. Now the question is whether we can also expect
solutions with separate vortices of the types $(1,0)$ and $(0,1)$.
Clearly if we put artificially $\gamma=0$ then eqs.(\ref{mcv125}) decouple
from each other and become similar to (\ref{mcv130}), so in this limit
there exist such isolated vortices. The opposite limit of $\alpha=\beta=0$
and $\gamma\neq 0$ characterises by existence of only hybrid $(1,1)$
vortices as physical excitations as was shown in \cite{only}.
In a more general case we have to
analyse carefully asymptotices of such potentially interesting solutions.
We substitute to Eqs.(\ref{mcv125}) following forms
$\rho_{I}=\rho_{I}^{0}(1+f_{I})$ and linearise them with respect to $f_{I}$'s
\begin{eqnarray}\label{mcv140}
  (\alpha\beta-\gamma^{2})\nabla^{2}f_{1}=
  2\alpha(\beta-\gamma)f_{1}+2\gamma(\alpha-\gamma)f_{2} \;,\nonumber\\
  (\alpha\beta-\gamma^{2})\nabla^{2}f_{2}=
  2\gamma(\beta-\gamma)f_{1}+2\beta(\alpha-\gamma)f_{2} \;,
\end{eqnarray}
Of course there is a solution with $f_{1}=f_{2}$ which corresponds to
the exact solution with only hybrid vortices. To look for a more
general solution let us take $f_{2}=f_{1}+u$:
\begin{eqnarray}\label{mcv150}
  \nabla^{2}u=\frac{2(\alpha-\gamma)(\beta-\gamma)}
                             {\alpha\beta-\gamma^{2}}u \;\;, \nonumber\\
  \nabla^{2}f_{1}-2f_{1}=\frac{2\gamma(\alpha-\gamma)}
                                         {\alpha\beta-\gamma^{2}}u \;\;,
\end{eqnarray}
The first equation is an eigenvalue problem which is solved by an
asymptotics of Bessel or modified Bessel function dependent on whether
the eigenvalue on R.H.S. is negative or positive respectively. Its sign is
the same as the sign of its denominator
$\alpha\beta-\gamma^{2}$. If this determinant is negative then
both $f_{1}$ and $f_{2}$ have an admixture of slowly vanishing
solution
\begin{equation}\label{mcv155}
   f_{1}=f - \frac{\alpha-\gamma}{\alpha+\beta-2\gamma} u \;\;,\;\;
   f_{2}=f + \frac{\beta-\gamma}{\alpha+\beta-2\gamma} u \;\;,
\end{equation}
where $f$ is a general exponentially vanishing solution of uniform
equation $\nabla^{2}f-2f=0$. The charge density fluctuations due to $u$
are
\begin{equation}\label{mcv156}
  \delta\rho_{1}=-\frac{1}{2}eB^{ext}
                  \frac{(\alpha-\gamma)(\beta-\gamma)}
                       {(\alpha\beta-\gamma^{2})(\alpha+\beta-2\gamma)} u
  \;\;\;,\;\;\;
  \delta\rho_{2}=-\delta\rho_{1}\;.
\end{equation}
A total electric charge of such a vortex does not have any contribution
from these fluctuations since they exactly cancel each other. Nevertheless
separate charges on each of the layers are divergent and not well
localised. This signals that if we take into account Coulomb interaction
such solutions even if substantially distorted can have much higher
energy then $(1,1)$ vortices.

For a positive determinant $u$ vanishes exponentially and there are no
problems with charge normalisation. Thus we can see that the line in the
parameter space $\alpha\beta=\gamma^{2}$ is a border between an area
where as physical excitations there are only hybrid $(1,1)$ vortices
($\alpha\beta<\gamma^{2}$) and that in which there is a possibility
of finite charge and energy separate $(1,0)$ and $(0,1)$ vortices
($\alpha\beta>\gamma^{2}$). In more physical terms if
the interaction $\gamma$ between the two layers is large enough
as compared to $\alpha$ and $\beta$ vortex can not form only on one of
the layers. Only the double-layer $(1,1)$ composites are allowed.

    Similar analysis can be performed at the bifurcation point
$\alpha=\beta=\gamma$. Eqs. (\ref{mcv150}) are replaced by
\begin{eqnarray}\label{mcv160}
  \nabla^{2}u=0                                          \;\;, \nonumber\\
  \nabla^{2}f_{1}-f_{1} = \frac{1}{2} u \;\;.
\end{eqnarray}
A general solution is given by
$f_{1,2}=f\stackrel{-}{+}\frac{1}{2}u_{0}$. $u_{0}$ is a constant
which is a parameter connected with the zero mode - it only changes
asymptotic values keeping $\rho_{1}+\rho_{2}$ fixed. This symmetry
is broken and the parameter is locked at $u_{0}=0$ if we take the Coulomb
interaction of the form (\ref{mcv123}). $f$ is an exponentially vanishing
asymptotics of modified Bessel function. Thus if $\alpha=\beta=\gamma$
and both condensates are nonvanishing the smallest quasiparticle
excitations are hybrid $(1,1)$ vortices. This perturbative result is in
agreement with conclusion in \cite[Eq.(2.19)]{semi} if we take into
account that in their paper boundary conditions are imposed in such a way
that one of the condensates vanishes. Our boundary
conditions can be obtained by global SU(2) transformation.

   Here we have just analysed properties of eventual single isolated
rotationally symmetric $(1,0)$ or $(0,1)$ vortex solution. Now we will try
to directly compare properties of $(1,1)$ vortex with a pair of
$(1,0)$ and $(0,1)$ vortices.

\section{ Splitting of hybrid vortex into separate
                                          $(1,0)$ and $(0,1)$ vortices }

  Keeping in mind the whole area $\alpha\beta<\gamma^{2}$ we will
restrict in this section for a sake of simplicity to the case
$\alpha=\beta=0$ and $\gamma>0$. Let the unperturbed hybrid vortex be
\begin{equation}\label{spl10}
  \psi=\phi=F(r)e^{i\theta} \;\;.
\end{equation}
Now we take perturbations of the phases of the scalar field to be
$\alpha_{1},\alpha_{2}$ and those of the moduli: $Fh_{1}$ , $Fh_{2}$.
\begin{eqnarray}\label{spl20}
  \psi+\delta\psi=F(r)[1+h_{1}(r,\theta)]
                                  e^{i\theta+i\alpha_{1}(r,\theta)} \;\;,
                                                                 \nonumber \\
  \phi+\delta\phi=F(r)[1+h_{2}(r,\theta)]
                                  e^{i\theta+i\alpha_{2}(r,\theta)} \;\;.
\end{eqnarray}
Linearisation of the self-dual equations (\ref{mcv80}) with respect
to perturbations of the scalar fields and those of gauge potentials
$c_{k}^{(I)}$ , $I=1,2$, yields
\begin{eqnarray}\label{spl30}
   c_{\theta}^{(I)}=\partial_{r}h_{I}-\frac{1}{r}\partial_{\theta}\alpha_{I}
                                                            \;\;, \nonumber \\
   c_{r}^{(I)}=-\frac{1}{r}\partial_{\theta}h_{I}-\partial_{r}\alpha_{I}
\end{eqnarray}
for $I=1,2$. Once the perturbations of the Higgs field are known,
$c_{k}^{(I)}$ can be calculated from the above equations. To have a unique
solution we have to fix the gauge
\begin{equation}\label{spl40}
  \partial_{k}c_{k}^{(I)}\equiv
    \partial_{r}c_{r}^{(I)}+\frac{1}{r}c_{r}^{(I)}+
    \frac{1}{r}\partial_{\theta}c_{\theta}^{(I)}=0 \;\;,\;\; I=1,2\;.
\end{equation}
Upon substitution of (\ref{spl30}) to this gauge condition we will obtain
\begin{equation}\label{spl50}
  \nabla^{2}\alpha_{1}=0 \;\;\;,\;\;\; \nabla^{2}\alpha_{2}=0 \;\;.
\end{equation}
Similar substitution of (\ref{spl30}) to Gauss' laws (\ref{mcv40})
will lead to
\begin{equation}\label{spl60}
  \nabla^{2}h_{1}=4\gamma F^{2}(r)h_{2} \;\;,\;\;
  \nabla^{2}h_{2}=4\gamma F^{2}(r)h_{1} \;\;.
\end{equation}
We would like to describe decay of the hybrid vortex in the direction
of $x$-axis on two separate vortices. That is the reason for symmetry
requirements on $h_{I}$
\begin{equation}\label{spl70}
  h_{1}(x,y)=h_{2}(-x,y) \;\;\;,\;\;\; h_{I}(x,y)=h_{I}(x,-y) \;\;.
\end{equation}
They restrict Fourier transforms with respect to $\theta$ to following
forms
\begin{eqnarray}\label{spl80}
  h_{1}(r,\theta)=H_{0}(r)+H_{1}(r)\cos\theta+\delta h_{1} \;,\nonumber\\
  h_{2}(r,\theta)=H_{0}(r)-H_{1}(r)\cos\theta+\delta h_{2} \;,
\end{eqnarray}
where $\delta h_{I}$ are tails of Fourier seria beginning at terms
$\cos 2\theta$ and $\sin 2\theta$. Substitution of these seria to
Eqs.(\ref{spl60}) leads to equations
\begin{eqnarray}\label{spl90}
  (\frac{d^{2}}{dr^{2}}+\frac{1}{r}\frac{d}{dr}) H_{0} =
                                              4\gamma H_{0} \;\;,\nonumber\\
  (\frac{d^{2}}{dr^{2}}+\frac{1}{r}\frac{d}{dr}-\frac{1}{r^{2}}) H_{1} =
                                             -4\gamma H_{1} \;\;.
\end{eqnarray}
Thus $H_{0}$ is an exponentially vanishing function while an asymptotics
of $H_{1}$ is equal to zero or proportional to the asymptotics
of Bessel function and slowly vanishing. But why should $H_{1}$ be nonzero?

    Its asymptotics near $r=0$ can be read from Eq.(\ref{spl90}), namely
$H_{1}(r)\sim\frac{-\xi}{r}$. With such an asymptotics
the scalar fields for small $r=0$ and small $\xi$ tend to
\begin{equation}\label{spl99}
  \psi\sim re^{i\theta}-\xi \;\;\;,\;\;\; \phi\sim re^{i\theta}+\xi \;\;,
\end{equation}
and zeros are shifted to $\xi$ and $-\xi$ respectively. Thus if $(1,0)$
and $(0,1)$ vortices are already not exactly on top of each other
then $\xi\neq 0$ and also $H_{(1)}$ has to be nonzero.

    Putting all the above together let us calculate changes of
electron numbers
\begin{equation}\label{spl100}
  \delta Q_{I}\equiv\int d^{2}x \; [ \; F^{2}(r)(1+h_{I})^{2}-F^{2}(r) \; ]
\;\;.
\end{equation}
With the forms (\ref{spl80}) the above integral can be finally rewritten as
\begin{equation}\label{spl110}
  \delta Q_{I} = \int d^{2}x\;F^{2}(r)[2H_{0}+H_{0}^{2}] +
                 \int d^{2}x\;F^{2}(r)[H_{1}^{2}+(\delta h_{I})^{2}] \;\;.
\end{equation}
The first integral is finite. To show that the second one is infinite
it is enough to calculate its first term
\begin{equation}\label{spl120}
  \int d^{2}x\;F^{2}(r)H_{1}^{2}(r)\sim const\;R  \;\;,
\end{equation}
where $R$ is a radial cut-off. Thus we can see that $\delta Q_{I}$'s
are badly divergent. This divergence is not an artifact of linearisations
we have done since it depends only on the asymptotics of $H_{(1)}$ for
large $r$ where the approximations are valid.

    In other words the lowest energy hybrid $(1,1)$ vortex can not decay
into separate $(1,0)$ and $(0,1)$ vortices because such a decay would
violate conservation of the numbers of particles. The decay could be
possible only from an excited state of hybrid vortex. A pair of separated
$(1,0)$ and $(0,1)$ vortices must be higher energy configuration than
a single hybrid vortex. Once again we are lead to conclusion that
hybrids will dominate statistics of the system.

\section{ Condensation of vortices and hierarchy of states }

   Let us consider the nondegenerate case $\alpha\beta-\gamma^{2}\neq0$.
We perform a singular gauge transformation
\begin{equation}\label{nc30}
   \omega_{I}\rightarrow\omega_{I}+\alpha_{I} \;\;,\;\;
   a_{\mu}^{(I)}\rightarrow a_{\mu}^{(I)}+\partial_{\mu}\alpha_{I} \;\;,
\end{equation}
where positions of vortices are encoded in singular shifts of phases
\begin{equation}\label{nc40}
   \alpha_{I}=\sum_{s_{I}} \Theta(z-z_{s_{I}}) \;\;,
\end{equation}
on the fields in the Lagrangian
\begin{eqnarray}\label{nc10}
  L&=&i\psi^{\star}D_{0}\psi-\frac{1}{2M}D_{k}\psi^{\star}D_{k}\psi
   +i\phi^{\star}D_{0}\phi-\frac{1}{2M}D_{k}\phi^{\star}D_{k}\phi
   -V(\psi,\phi)                                               \nonumber \\
   &-&\frac{1}{4X}\varepsilon^{\mu\nu\lambda}
                  a_{\mu}^{(1)}\partial_{\nu}a_{\lambda}^{(1)}
     -\frac{1}{4Y}\varepsilon^{\mu\nu\lambda}
                  a_{\mu}^{(2)}\partial_{\nu}a_{\lambda}^{(2)} \nonumber \\
   &-&\frac{1}{4Z}\varepsilon^{\mu\nu\lambda}
                  a_{\mu}^{(1)}\partial_{\nu}a_{\lambda}^{(2)}
     -\frac{1}{4Z}\varepsilon^{\mu\nu\lambda}
                  a_{\mu}^{(2)}\partial_{\nu}a_{\lambda}^{(1)}-V  \;\;,
\end{eqnarray}
Under such a transformation which is intented as an introduction of new
vortex degrees of freedom the Lagrangian changes by
\begin{eqnarray}\label{nc50}
  \triangle L_{vortex}&=&-\pi( \frac{a_{\mu}^{(1)}K^{\mu}_{1}}{X}
  + \frac{a_{\mu}^{(2)}K^{\mu}_{2}}{Y}
  + \frac{ a_{\mu}^{(2)}K^{\mu}_{1} + a_{\mu}^{(1)}K^{\mu}_{2} }{Z})
                                                          \nonumber \\
  &-&\frac{1}{2}\pi( \frac{\partial_{\mu}\alpha_{1}K^{\mu}_{1}}{X}
                   + \frac{\partial_{\mu}\alpha_{1}K^{\mu}_{1}}{Y}
                   + \frac{\partial_{\mu}\alpha_{1}K^{\mu}_{2}
                         +\partial_{\mu}\alpha_{2}K^{\mu}_{1}}{Z} ) \;\;,
\end{eqnarray}
where vortex currents are
\begin{equation}\label{nc60}
  K^{\mu}_{I}=\frac{1}{2\pi}\varepsilon^{\mu\nu\lambda}
              \partial_{\nu}\partial_{\lambda}\alpha_{I}=
              \sum_{s_{I}}\dot{x}^{\mu}_{s_{I}}\delta^{(2)}(z-z_{s_{I}}) \;\;,
\end{equation}
with the parametrisation of vortex positions by
$x^{\mu}_{s_{I}}=(t,x^{k}_{s_{I}})$. With this form of the vortex current
one can easily rewrite the change in the Lagrangian (\ref{nc50}) as
\begin{eqnarray}\label{nc70}
  \triangle L_{vortex}=-\pi( \frac{a_{\mu}^{(1)}K^{\mu}_{1}}{X}
  + \frac{a_{\mu}^{(2)}K^{\mu}_{2}}{Y}
  + \frac{ a_{\mu}^{(2)}K^{\mu}_{1} + a_{\mu}^{(1)}K^{\mu}_{2} }{Z}) &
                                                          \nonumber \\
   - \pi[ \frac{1}{X}\sum_{p_{1}>s_{1}}\dot{\Theta}(z_{p_{1}}-z_{s_{1}})
        + \frac{1}{Y}\sum_{p_{2}>s_{2}}\dot{\Theta}(z_{p_{2}}-z_{s_{2}})
        + &\frac{1}{Z}\sum_{s_{1},s_{2}}\dot{\Theta}(z_{s_{1}}-z_{s_{2}}) ]
                                                                      \;\;,
\end{eqnarray}
 The charges of vortices with respect to the C-S fields can be read
from constraint equations
\begin{eqnarray}\label{nc80}
  \triangle\rho_{1}=-\frac{1}{2X}\triangle B_{1}
                           -\frac{1}{2Z}\triangle B_{2} \;\;, \nonumber \\
  \triangle\rho_{2}=-\frac{1}{2Z}\triangle B_{1}
                           -\frac{1}{2Y}\triangle B_{2} \;\;,
\end{eqnarray}
where by $\triangle$ we denote deviations with respect to uniform
background. Thus upon integrations and use of quantisation conditions
we find that charges of vortices with winding numbers $(p,q)$, if they
exist, are given by
\begin{eqnarray}\label{nc90}
  q_{1}[p,q]=\pi\frac{\beta p-\gamma q}{\alpha\beta-\gamma^{2}}
                                                          \;\;, \nonumber \\
  q_{2}[p,q]=\pi\frac{\alpha q-\gamma p}{\alpha\beta-\gamma^{2}}
                                                          \;\;,
\end{eqnarray}
while their electric charges are
\begin{equation}\label{nc100}
  eQ[p,q]=-e(q_{1}+q_{2})=
  -e\pi\frac{(\beta-\gamma)p+(\alpha-\gamma)q}{\alpha\beta-\gamma^{2}} \;\;.
\end{equation}

  Let us concentrate on the case $\alpha\beta-\gamma^{2}<0$ in which
dominating vortices are
those with winding numbers (1,1) - the hybrid anyons. The vortex currents
can be identified $K^{\mu}_{1}=K^{\mu}_{2}=K^{\mu}$, the charges
$q_{1}=\pi\frac{\beta-\gamma}{\alpha\beta-\gamma^{2}}$ and
$q_{2}=\pi\frac{\alpha-\gamma}{\alpha\beta-\gamma^{2}}$ are positively
definite. The vortex part of the Lagrangean can be rewritten as
\begin{equation}\label{nc110}
  \triangle L_{vortex}=
      -q_{1}a_{\mu}^{(1)}K^{\mu}-q_{2}a_{\mu}^{(2)}K^{\mu}
      +\frac{\hat{\kappa}}{\pi}\sum_{p<q}\dot{\Theta}(z_{p}-z_{q}) \;\;,
\end{equation}
where the statistical parameter $\hat{\kappa}$ is defined modulo integer
multiplicity of $2\pi$
\begin{equation}\label{nc120}
  \hat{\kappa}=-\pi^{2}(\frac{\alpha+\beta-2\gamma}{\alpha\beta-\gamma^{2}})
                                                            -2\pi q \;\;.
\end{equation}
Upon variation of the effective Lagrangian with respect to $a_{0}^{(I)}$
we get a new set of constraint equations
\begin{eqnarray}\label{nc130}
\rho_{1}&=&-\frac{1}{2X}B_{1}-\frac{1}{2Z}B_{2}+q_{1}\rho_{vortex}\;,\nonumber\\
\rho_{2}&=&-\frac{1}{2Z}B_{1}-\frac{1}{2Y}B_{2}+q_{2}\rho_{vortex}\;.
\end{eqnarray}
With the identification $K^{0}=\rho_{vortex}$ these equations are to be
understood in terms of mean values. For large enough separations of
vortices the Lagrangian (\ref{nc110}) can be supplemented by kinetic
term similarly as in \cite{ei2}
\begin{equation}\label{nc140}
  \triangle H_{vortex}=\frac{1}{2M_{v}}\sum_{p}
  [p^{k}_{p}-q_{1}a_{k}^{(1)}-q_{2}a_{k}^{(2)}+c_{k}]^{2} \;\;.
\end{equation}
The auxillary vector field is given by
$c_{k}(z)=\frac{\hat{\kappa}}{\pi}\sum_{p}\partial_{k}\Theta(z-z_{p})$ or
by a constraint
\begin{equation}\label{nc150}
  \varepsilon_{mn}\partial_{m}c_{n}=2\hat{\kappa}\rho_{vortex} \;\;.
\end{equation}
It is worthwhile to mention that the form of the Hamiltonian (\ref{nc140})
is justified only if separations of vortices are large as compared
to sizes of their cores. In the derivarion vortices were represented
by infinitely thin delta-like distributions of magnetic flux and charge.
Short range interactions of hybrid $(1,1)$ anyons were analysed
in \cite{only} in the case of only mutual interactions. The conclusions
can be easily generalised to this case. The main result is that statistical
interactions mediated in (\ref{nc140}) by the fields $c_{k}$ are effectively
switched off at short distances. Instead there are short range
charge-flux interactions as if vortices were finite-width charged selenoids.
They can lead to magnetic trapping and quantum mechanically to scattering
resonances characterised by a set of Landau levels.

    The Hamiltonian (\ref{nc140}) leads to the following field-theoretical
description
\begin{equation}\label{nc160}
  H_{vortex}=\frac{1}{2M_{v}}
  \mid (\partial_{k}-iq_{1}a_{k}^{(1)}-iq_{2}a_{k}^{(2)}+ic_{k})u\mid^{2}\;\;,
\end{equation}
The self-dual equation for vortex field $D_{+}u=0$ in a uniform state
implies that $q_{1}B_{1}+q_{2}B_{2}+\varepsilon_{mn}\partial_{m}c_{n}=0$.
This and the relations $B_{1}=B_{2}=eB_{ext}$ lead to the filling factor
for vortices
\begin{equation}\label{nc170}
  \nu_{vortex}\equiv\frac{2\pi\rho_{vortex}}{QB_{ext}}=\frac{\pi}{\hat{\kappa}}
\;.
\end{equation}
{}From equations (\ref{nc130}) we get the electronic filling factor
\begin{equation}\label{nc180}
  \nu^{(1)}=\frac{1}{\xi+\frac{1}{2q}} \;\;\;,\;\;\;
                                             \xi=\frac{1}{\nu^{(0)}} \;\;.
\end{equation}
This formula is the same as in the single layer quantised Hall effect
\cite{laughlin}. As such it has to be compared with that for the double-layer
effect when there are single vortices of the types $(1,0)$ and $(0,1)$.
Such a formula was derived in the article by Ezawa and Iwazaki \cite{ei2}.
\begin{equation}\label{nc190}
  \nu^{(1)}=\frac{1}{\xi+\frac{1}{4(p+r)}}=
                                       \frac{1}{\xi+\frac{1}{4q}} \;\;,
\end{equation}
where $p,r$ are integers and can be replaced by $q=p+r$. The characteristic
difference between formulas (\ref{nc180}) and (\ref{nc190}) is the
multiplyer 2 instead of 4 when there is dominance by hybrid vortices
$(1,1)$.

\section{Conclusions}

    There are two characteristic areas in the parameter space.
If $\alpha\beta>\gamma^{2}$ vortices with winding numbers $(1,0)$ and $(0,1)$
dominate while for $\alpha\beta<\gamma^{2}$ quite as well as for the
bifurcation point $\alpha=\beta=\gamma$ the most important type are
$(1,1)$ hybrids. This dominance by one type or the other can be observed
thanks to characteristic changes in Haldane-Halderin hierarchy of states.

    The hybrid $(1,1)$ vortices can be thought of as composites of
separate $(1,0)$ and $(0,1)$ vortices. There is always present charge-flux
interaction between them being a remnant of long range statistical
Aharonov-Bohm-type interaction. This interaction acts also between
whole hybrid vortices and can lead to magnetic trapping and
at the quantum level to scattering resonances characterised by Landau levels.
If $\alpha\beta<\gamma^{2}$ then $(1,0)$ and $(0,1)$ vortices are additionally
attracted by potential force to form a hybrid vortex.

     The distribution of positively charged background may be important
for the range of physical states. If the background is the same for the two
layers the Coulomb interaction will allow only states of the type $(k,k,m)$
with two charge densities equal one another.

\end{document}